\documentclass[]{spie} 
\usepackage[dvips]{graphicx} 

\title{TRIDENT: an infrared camera optimized for the detection of methanated substellar companions around nearby stars}

\author{Christian Marois, Ren\'{e} Doyon, Daniel Nadeau, Ren\'{e} Racine,\\ Martin Riopel, Philippe Vall\'{e}e
\skiplinehalf
D\'{e}partement de physique, Universit\'{e} de Montr\'{e}al, C.P. 6128,\\ Succ. A, Montr\'{e}al, QC, Canada H3C 3J7
}

\begin{document}  
  \maketitle  

\begin{abstract} 
A near-infrared (0.85-2.5 $\mu$m) camera in use at the Canada-France-Hawaii Telescope and at the 1.6m telescope of the Observatoire du Mont-M\'{e}gantic is described. The camera is based on a Hawaii-1 1024$\times$1024 HgCdTe array detector. Its main feature is to acquire three simultaneous images at three wavelengths (simultaneous differential imaging) across the methane absorption bandhead at $1.6\mu$m, enabling an accurate subtraction of the stellar point spread function (PSF) and the detection of faint close methanated companions. The instrument has no coronagraph and features a fast (1 MHz) data acquisition system without reset anomaly, yielding high observing efficiencies on bright stars. The performance of the instrument is described, and it is illustrated by CFHT images of the nearby star $\upsilon$ And. TRIDENT can detect ($3\sigma$) a methanated companion with $\Delta H$=10 at $0.5^{\prime \prime}$ from the star in one hour of observing time. Non-common path aberrations between the three optical paths are the limiting factors preventing further PSF attenuation. Reference star subtraction and instrument rotation improve the detection limit by one order of magnitude.
\end{abstract} 



\section{INTRODUCTION}
This paper describes a near-infrared camera built at the Laboratoire d'Astrophysique Exp\'{e}rimentale associated with the Observatoire du Mont-M\'{e}gantic (OMM) for use either on the 3.6m Canada-France-Hawaii telescope (CFHT) or the 1.6m telescope of the OMM. The goal was to build a compact simple instrument that would be easy to carry from the laboratory to any observatory. Commercial components were used as much as possible to minimize development time. 

The camera, called TRIDENT, is based on a Hawaii-1 1024$\times$1024 HgCdTe array detector sensitive from 0.85 to 2.5~$\mu$m. The main scientific objective of the camera is to search for faint companions (brown dwarfs or planets) around nearby stars. Ground-based detection of faint companions is difficult because of the atmospheric turbulence that distorts the stellar diffraction pattern and optical aberrations that produce quasi-static PSF structures. TRIDENT is designed to operate with an adaptive optics (AO) system (PUEO at CFHT, \cite{Rigaut98} or the OMM AO system\cite{Nadeau2002,Liviu2002}) to benefit from diffraction-limited images. The camera features a special beam-splitter that allows the acquisition of three simultaneous images in three distinct narrow spectral bands across the methane absorption bandhead at $1.6\mu$m, making possible very good subtraction of atmospheric speckles and quasi-static aberrations in the telescope and the AO system. If the three simultaneous wavelengths are carefully chosen, it is possible to enhance the star/companion contrast after image combination by selecting a spectral feature observed only in the companion but not in the star \cite{Smith87,Rosen96}. In TRIDENT, the three wavelengths (1.567~$\mu$m, 1.625~$\mu$m and  1.680~$\mu$m, 1\% bandwidth) have been selected across the 1.6 $\mu$m methane absorption bandhead, a spectral feature observed only in cold ($T_{\rm{eff}}$ $<$ 1500~K \cite{Fegley1996}) substellar objects such as T-type brown dwarfs and giant planets. TRIDENT is the practical realization of the simultaneous differential imaging concept\cite{Racine99,Marois2000}. First light of the instrument was achieved at the Observatoire du mont M\'{e}gantic in April 2001 (see figure 1).

\begin{figure}[h]
 \begin{center}
 \includegraphics[height=13cm]{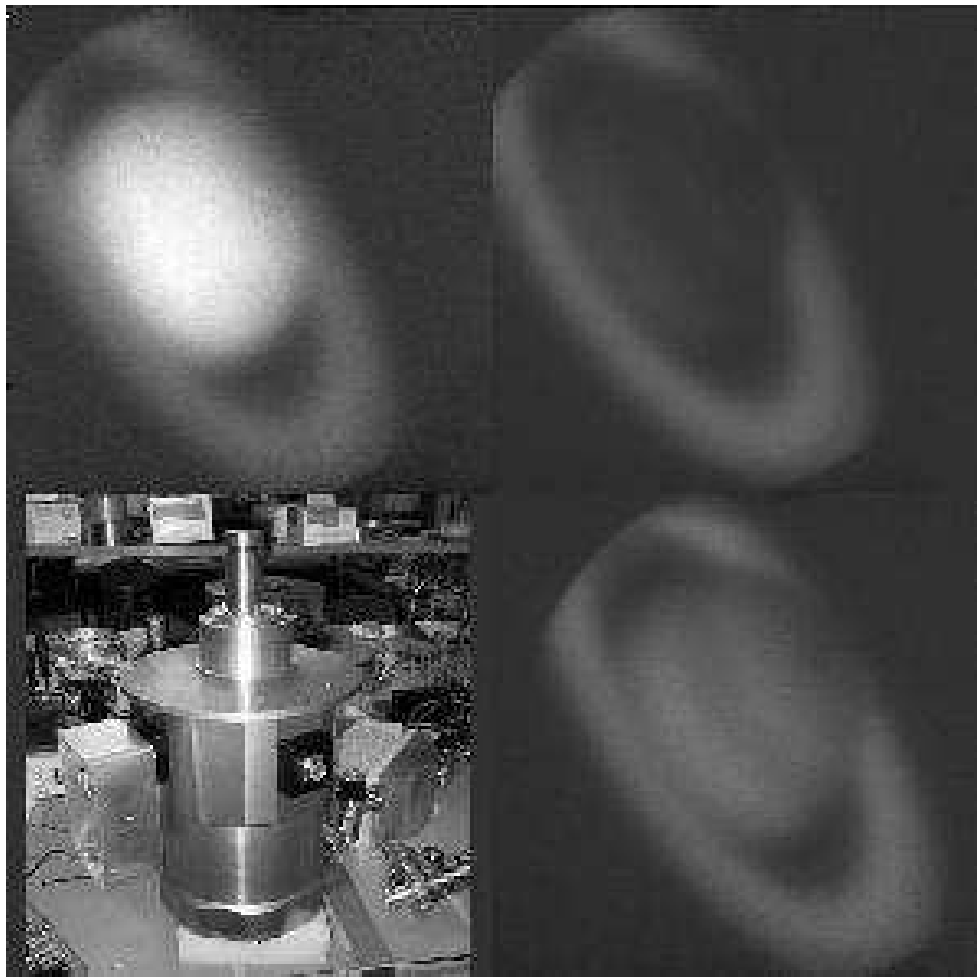}
 \caption{TRIDENT first light of Saturn at the Observatoire du Mont M\'{e}gantic in April 2001. Upper left corner is the 1.568~$\mu $m image, upper right is the 1.680~$\mu $m image, lower right corner is the 1.625~$\mu $m image and lower left is a picture of the instrument cryostat. The continuum light scattered by Saturn's ring appears equally bright in the three images while the planet is bright in the 1.568~$\mu $m image but faint in the other two images, due to methane absorption in Saturn's atmosphere.}
 \end{center}
\end{figure}

\section{Optical Design}

\begin{figure}[h]
 \begin{center}
\includegraphics[height=13cm]{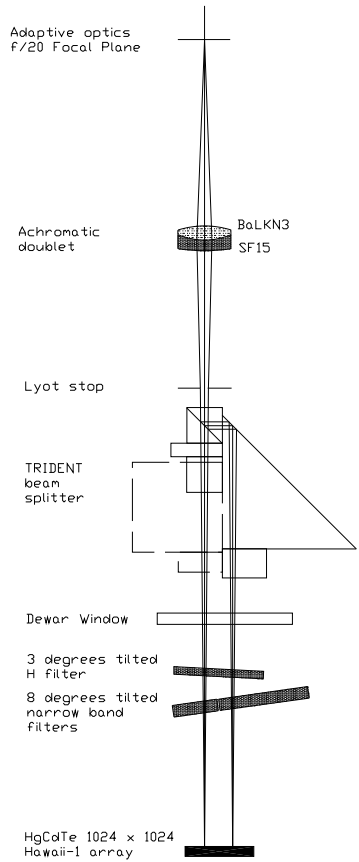}
 \caption{Optical layout of TRIDENT. The distance from the adaptive optics focal plane to the detector surface is 228mm. The optical paths for two of the three simultaneous images are seen in the representation. The third one is in a perpendicular view not visible in the figure. See text for a detailed description.}
 \end{center}
\end{figure}

An optical layout of the instrument is shown in Figure 2. The optical design is split in two sections: i) a field stop, achromatic doublet lens, Lyot stop, beam splitter and cryostat window all at room temperature and ii) an $H$-band filter to block thermal radiation, three narrow-band filters and the detector inside the cryostat. The diameter of the circular field stop is chosen to obtain the maximum field of view (FOV) with minimum overlap. A commercial achromatic doublet lens (Melles Griot Lal011) is used to re-image the focal plane on the detector and to create an image of the pupil on the Lyot stop. The image scale is $0.018^{\prime \prime}$ pixel$^{-1}$ at CFHT and $0.036^{\prime \prime}$ pixel$^{-1}$ at OMM, corresponding to 5 pixels/FWHM. The main reason for such an oversampling is to ensure accurate PSF rescaling and registering and avoid interpolation noise that could mask faint companions. The same Lyot stop can be used at both the CFHT and OMM as both telescopes have the same f/8.0 focal ratio. The 2~mm Lyot stop is used to reduce scattered light. The beam splitter consists of a combination of polarizing beam splitters, a wavelength retarder and right angle prisms. First, the light is divided by a polarizing beam splitter. One beam is reflected to a prism and then reflected to the detector. The second beam passes through a first order quarter wavelength retarder (optimized at $1.6 \mu $m) to transform the beam from linear to circular polarization. A second beam splitter is used to divide this beam in two. One beam goes directly to the detector to form the second channel while the second one is reflected to a second prism to be redirected to the detector. The beam splitter forms three simultaneous channels organized in a ``L'' shape on the detector. Each channel is centered on one quadrant of the detector so that all three images can be read simultaneously. Thin sheets of glass were optically cemented under the two prisms to ensure co-focality of the three channels. All optical components of the beam splitter were optically cemented together using a Norland optical Adhesive with UV curing.

Light enters the cryostat through a BK7 window located just below the beam splitter followed by a cold $H$-band filter tilted by 3 degrees used for blocking thermal radiation as the narrow-band filters are not blocked beyond 1.8 $\mu$m. Finally, three narrow-band (1\%) filters tilted by 8 degrees intercept the three beams before they reach the detector. All filters were manufactured by Barr Associates. Tilting the filters makes sure that most of the ghosts fall outside the FOV, which is critical for accurate relative registration and scaling of the three channel PSFs. Ghosts due to internal reflections within the two-substrate configuration of the narrow-band filters were unavoidable however. 

\begin{figure}[h]
 \begin{center}
 \includegraphics[width=17cm]{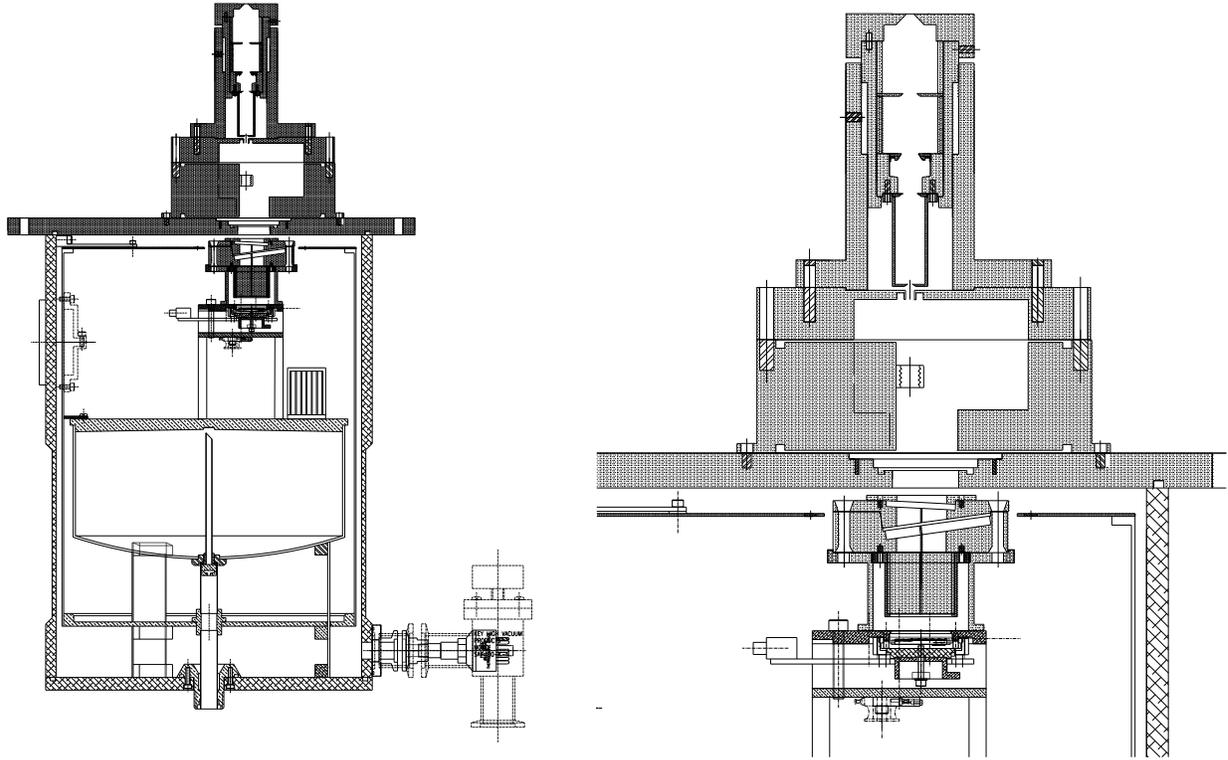}
 \caption{TRIDENT projected side view. Left: global view of the instrument. Right: details of the opto-mechanical bench.}
 \end{center}
\end{figure}

\section{Mechanical Design}

A mechanical layout of the cryostat is shown in Figure 3. The camera is housed in a ND8 Infrared Laboratories (Tucson, Az) cryostat with a diameter of 21~cm and an overall height of 32~cm (see Fig. 1). It contains a single liquid nitrogen reservoir with a hold time of 50~h. A temperature controller (Lakeshore model 331) is used to stabilize the detector temperature at $80$~K within mK precision. There is no moving part. Except for the cryostat, the opto-mechanical bench components were designed and built in-house. 

\section{Data Acquisition System}

The detector is a Hawaii-1 1024$\times$1024 HgCdTe array, mounted on a IRlab fanout board. The four video channels, one for every quadrant, are first pre-amplified in parallel just outside the cryostat and then fed to a SDSU-2 controller where the analog-to-digital conversion takes place. The measured read noise is 18 e$^{-}$. The minimum conversion time of 1~$\mu$s is used for reading the array as fast as possible to allow unsaturated observations of bright stars with high efficiency. The frame rate is 0.262~s. On board image co-addition enables Fowler sampling \cite{Fowler1990} to minimize readout noise. The system also features a unique clocking pattern that eliminates the reset anomaly of the Hawaii array without loss of observing time \cite{Riopel2002}. These features make the data-acquisition process very efficient and simple.

The host computer is a SUN microsystem Ultra 5 with a PCI acquisition board connected to the SDSU-2 via a 40 Mbits/sec fiber optic link for fast data transfer (0.75s per 1024$\times$1024 pixels, 32 bit image).

\section{Data-Acquisition Software}
During an observing session, all the control is done from a single computer display. A C-based program with Tcl/Tk user interface with a shared memory link to DS9 is controlling image acquisition and display. The data are stored on disk as 32 bit integers in FITS format \cite{Wells1981} with a header that includes all the observational parameters. The computer can be used at any time to look at recently acquired images to ascertain the data quality with the DS9 display program. An example of a TRIDENT image is shown in figure 4. Images can be downloaded to another computer in the local network to do further image analysis. An IDL-based software reduction pipeline is used to do preliminary image reduction and analysis (dark subtraction, flat field normalization, bad pixel corrections and image trimming, registration, scaling and subtraction) while observing at the telescope. A second IDL program iterates to find optimal PSF transformation parameters.

\begin{figure}[h]
 \begin{center}
 \includegraphics[width=12.5cm]{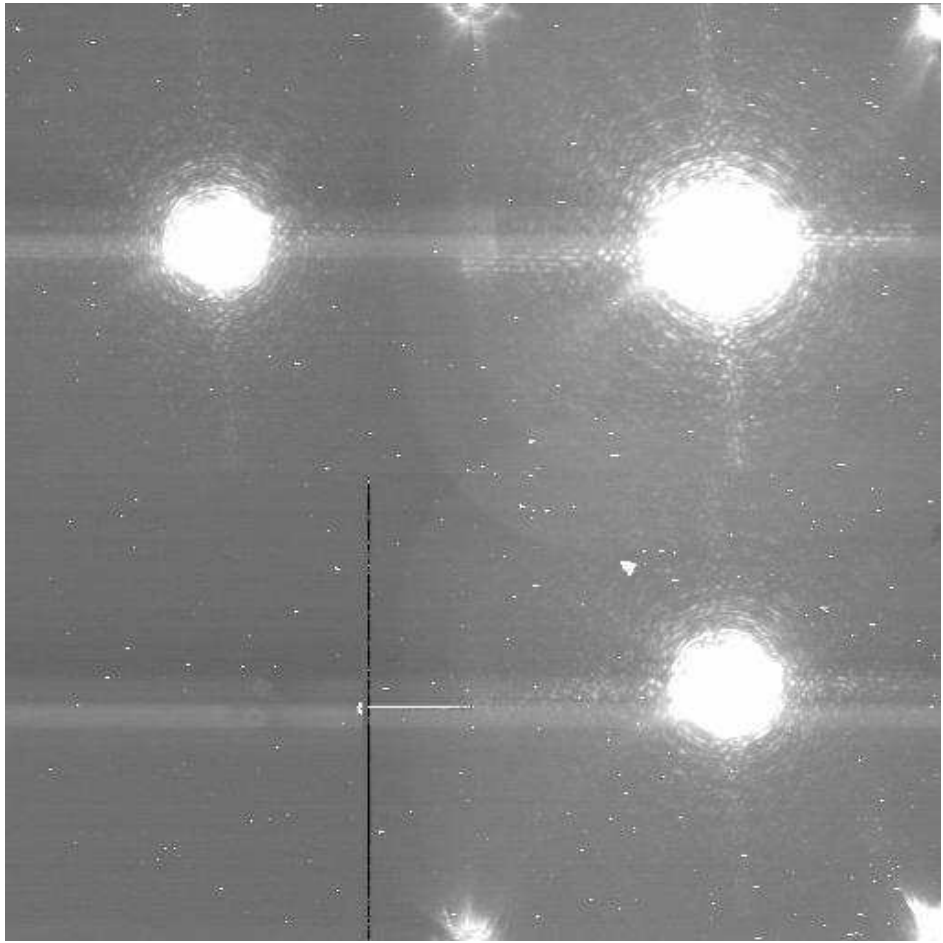}
 \caption{TRIDENT typical unprocessed image obtained at CFHT with the PUEO AOB. The upper left corner is the 1.568~$\mu $m image, upper right is the 1.680~$\mu $m image and lower right is the 1.625~$\mu $m image. Cross talk (horizontal bright features) is visible between quadrants. The FOV is $9^{\prime \prime}$ per quadrant at CFHT and $18^{\prime \prime}$ at OMM. Amplifier glow is seen in the lower and upper right quadrant corners.}
 \end{center}
\end{figure}

\section{Performance}
The TRIDENT performance is illustrated by data obtained at CFHT during the night of 2001 November 17, on the star $\upsilon $ And (see Marois et al.\cite{Marois2002} for more details on TRIDENT observations and Marois et al.\cite{Marois2000} for image combination algorithms). Figure 5 shows the $3\sigma $ detection threshold. Non-common path aberrations are slightly decorrelating the three PSFs, making impossible a perfect stellar PSF subtraction. To improve the instrument performance, a reference star was acquired and instrument rotations were applied. These two techniques improve the detection limit by one order of magnitude. To increase PSF correlation, and thus subtraction performance, a new optical system is presently under construction consisting of a better beam splitter and filters placed near the detector to minimize wavefront degradation. The goal is photon noise limited performance for long integrations (1h) without the need of a reference star or instrument rotation for a 3rd magnitude star at CFHT with the PUEO AO system. Predicted performance ($3\sigma $) is $\Delta H = 13$ at $0.5^{\prime \prime}$ separation.

\begin{figure}[h]
 \begin{center}
 \includegraphics[width=15cm]{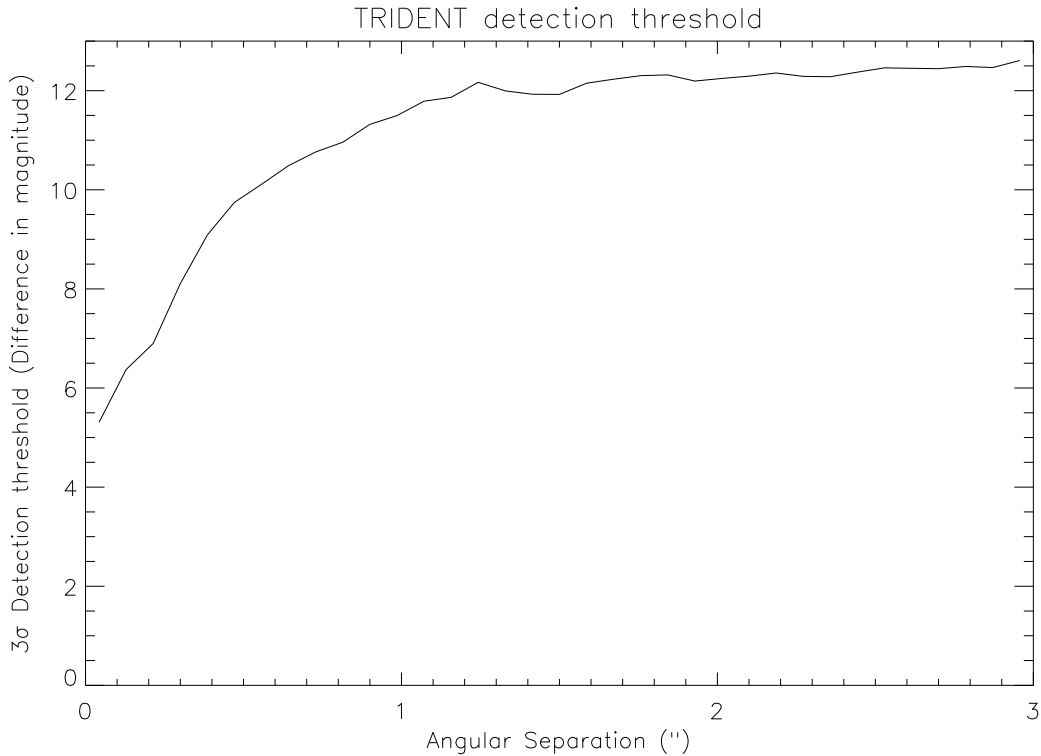}
 \caption{TRIDENT $3\sigma $ detection threshold for differential simultaneous imaging with reference star subtraction from November 2001 CFHT $\upsilon $ And (Object) and $\chi $ And (reference) data sets (1h integration per target). A 10 magnitude fainter companion would be detected $0.5^{\prime \prime}$ away from the primary.}
 \end{center}
\end{figure}

At present, TRIDENT does not make use of a coronagraph. A coronagraph increases detection efficiency by suppressing coherent photons and their associated photon noise from the primary star PSF. In practice, TRIDENT is not limited by coherent photon noise for integration time less than or equal to a minute, but by quasi-static and non-common path aberrations. The coronagraph would thus not increase the TRIDENT detection threshold, but only reduce the time needed to be dominated by quasi-static and non-common path aberrations. The typical Strehl ratio of 0.5 obtained with the PUEO adaptive optics system in the $H$ band would also lower coronograph efficiency since half the flux goes to a non-coherent halo. The use of a coronagraph with differential simultaneous imaging will be considered when adaptive optic systems provide higher Strehl ratios. Differential atmospheric refraction (as shown in figure 6) and optical flexions will have to be corrected to assure proper alignment between the star PSF and the coronagraph occulting mask at all observing wavelengths.
\clearpage
\begin{figure}[h]
 \begin{center}
 \includegraphics[width=12cm]{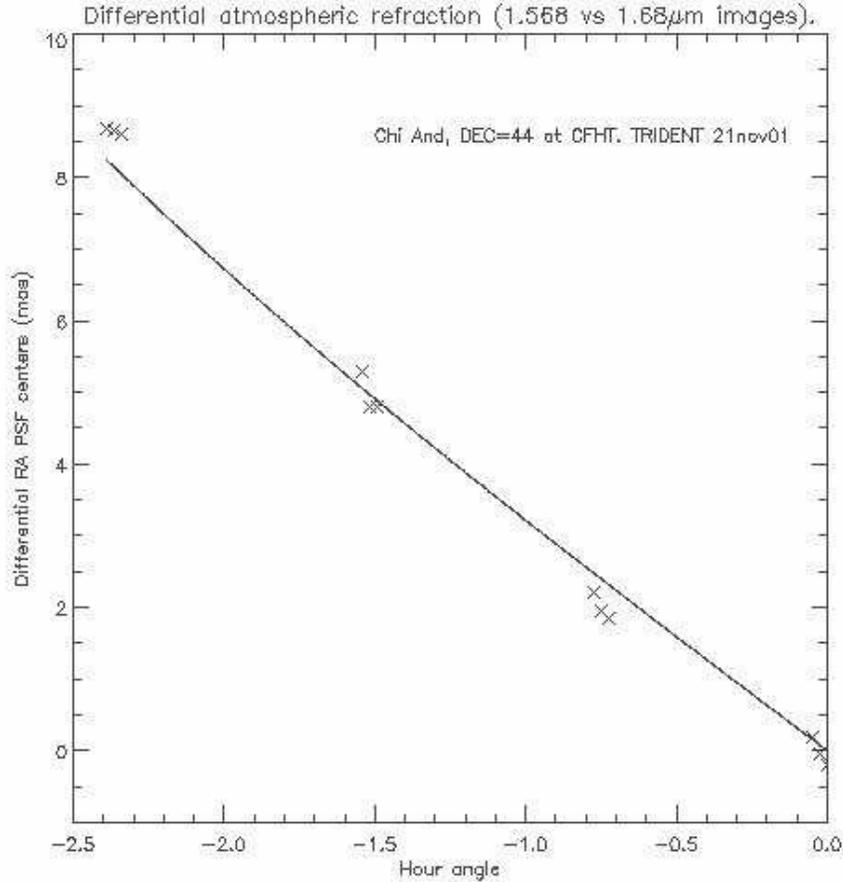}
 \caption{Differential atmospheric refraction between 1.568~$\mu $m and 1.680~$\mu $m images at CFHT with PUEO adaptive optics. Typical PSF differential center drift was 8 mas for 2.5 hours of integration time on $\chi$ And. The solid line represents a fitted differential refraction model from Edl\'{e}n\cite{Edlen1953}.}
 \end{center}
\end{figure}

\section{Conclusion}
The TRIDENT simultaneous multi-wavelength imaging technique is a powerful way to subtract a star aberrated PSF to search for nearby faint companions. Detection thresholds are as good as or better than what has been achieved on any other telescope, showing the potential of the observing technique. Non-common path aberrations need to be kept to a minimum to assure good correlation between the three simultaneous images and achieve good subtraction performance. A new optical design is presently being integrated and will enable an improvement of the PSF subtraction by at least one order of magnitude.

\acknowledgments     
 
This work is supported in part through grants from the Natural Sciences and Engineering Research Council, Canada and from Fonds FQRNT, Qu\'{e}bec.
\clearpage
\bibliography{art3spieRD}   

\begin{thebibliography}{10}

\bibitem{Rigaut98}
F.~{Rigaut}, D.~{Salmon}, R.~{Arsenault}, J.~{Thomas}, O.~{Lai}, D.~{Rouan},
  J.~P. {V{\' e}ran}, P.~{Gigan}, D.~{Crampton}, J.~M. {Fletcher},
  J.~{Stilburn}, C.~{Boyer}, and P.~{Jagourel}, ``{Performance of the
  Canada-France-Hawaii Telescope Adaptive Optics Bonnette},'' {\em PASP} {\bf
  110}, p.~152, 1998.

\bibitem{Nadeau2002}
D.~{Nadeau}, L.~{Ivanescu}, and R.~{Racine}, ``{A Simplified Adaptive Optics
  System},'' in {\em Adaptive Optical System Technologies II, Proc. SPIE},
  P.~L. Wizinowich and D.~Bonaccini, eds.,  {\bf 4839}, Bellingham, 2002.

\bibitem{Liviu2002}
L.~{Ivanescu}. {PhD. Thesis}, 2002.

\bibitem{Smith87}
W.~H. {Smith}, ``{Spectral differential imaging detection of planets about
  nearby stars},'' {\em PASP} {\bf 99}, p.~1344, 1987.

\bibitem{Rosen96}
E.~D. {Rosenthal}, M.~A. {Gurwell}, and P.~T.~P. {Ho}, ``{Efficient Detection
  of Brown Dwarfs Using Methane-Band Imaging.},'' {\em NATURE} {\bf 384},
  p.~243, 1996.

\bibitem{Fegley1996}
B.~J. {Fegley} and K.~{Lodders}, ``{Atmospheric Chemistry of the Brown Dwarf
  Gliese 229B: Thermochemical Equilibrium Predictions},'' {\em ApJL} {\bf 472},
  p.~L37, 1996.

\bibitem{Racine99}
R.~{Racine}, G.~{Walker}, D.~{Nadeau}, R.~. {Doyon}, and C.~{Marois},
  ``{Speckle Noise and the Detection of Faint Companions},'' {\em PASP} {\bf
  111}, p.~587, 1999.

\bibitem{Marois2000}
C.~{Marois}, R.~{Doyon}, R.~{Racine}, and D.~{Nadeau}, ``{Efficient Speckle
  Noise Attenuation in Faint Companion Imaging},'' {\em PASP} {\bf 112}, p.~91,
  2000.

\bibitem{Fowler1990}
A.~M. {Fowler} and I.~{Gatley}, ``{Demonstration of an algorithm for read-noise
  reduction in infrared arrays},'' {\em ApJL} {\bf 353}, p.~L33, 1990.

\bibitem{Riopel2002}
M.~{Riopel}, R.~{Doyon}, D.~{Nadeau}, and C.~{Marois}, ``{An Optimized Data
  Acquisition System Without Reset Anomaly for the Hawaii and Hawaii-2
  arrays},'' in {\em {Scientific Detectors for Astronomy Workshop, Waimea,
  Hawaii}},  2002.

\bibitem{Wells1981}
D.~C. {Wells}, E.~W. {Greisen}, and R.~H. {Harten}, ``{FITS - a Flexible Image
  Transport System},'' {\em AAPS} {\bf 44}, p.~363, 1981.

\bibitem{Marois2002}
C.~{Marois}, D.~{Nadeau}, R.~{Doyon}, and R.~{Racine}, ``{Differential
  Simultaneous Imaging and Faint Companions: First Results from CFHT},'' in
  {\em Brown Dwarfs: Proceedings of IAU Symposium 211},  E.~L. Mart\'{\i}n,
  ed.,  {\bf 211}, 2003.

\bibitem{Edlen1953}
B.~{Edl\'{e}n} {\em J. Opt. Soc. Amer.} {\bf 43}, p.~339, 1953.

\end{thebibliography}
\bibliographystyle{spiebib}   

\end{document}